\documentstyle[12pt]{article}
\newcommand{\be}{\begin{equation}}
\newcommand{\ee}{\end{equation}}
\newcommand{\bi}[1]{\vspace{-3mm} \bibitem{#1}}

\begin{document}

\centerline{\large \it Physics Letters B \ 323 (1994) 296-304.}
\vskip 11mm
\centerline{\Large \bf Bosonic String in 
Affine-Metric Curved Space.} 
\vskip 7 mm 
\centerline{\large Vasily E. Tarasov } 
\vskip 5 mm
\centerline{Theoretical High Energy Physics Dep.} 
\vskip 1 mm
\centerline{Nuclear Physics Institute, Moscow State
University, 119992 Moscow, Russia} 
\vskip 1 mm
\centerline{\small E-mail: TARASOV@THEORY.SINP.MSU.RU} 
\vskip 11 mm
The sigma model approach to the closed bosonic string on the
affine-metric manifold is considered. 
The two-loop metric counterterms for the nonlinear 
two-dimensional sigma model with affine-metric target
manifold are calculated. 
The correlation of the metric and affine connection 
is considered as the result of the conformal invariance 
condition for the nonlinear sigma model.
The examples of the nonflat nonRiemannian manifolds 
resulting in the trivial metric beta-function are suggested.
\vskip 5 mm

\section{Introduction.}

String theory in a curved space is a consistent quantum theory if 
the quantum nonlinear two-dimensional sigma model \cite{Frid,Alv} 
is conformally invariant. The conformal invariance
requires that the sigma model beta-functions  \cite{Frid,Alv} 
be trivial \cite{Shor}. Since the conformal
anomaly of the nonlinear sigma model depends on the geometrical structures 
(on the background fields) of the curved space (manifold), the 
beta-function vanishing condition lead to the restrictions on consistent
structures (backgrounds fields).

Different geometrical structures can be defined on the manifold
\cite{Vas}. In the bosonic case a metric and a connection
structures are used. Riemannian manifolds are considered as 
a field  manifold for usual nonlinear sigma-model \cite{Frid,Alv}.
The connection structure of this manifold is uniquely constructed 
from metric, i.e. the "strong" correlation between the connection and
metric structures is postulated. In the general case, 
these structures are not correlated \cite{Sch} and the curved space
is nonmetric nonRiemannian manifold.  Therefore it was
suggested to consider the nonlinear sigma model with nonmetric (affine-metric)
manifold \cite{Tarb1} and to obtain the correlation between the
metric and connection structures as the result of
the ultraviolet finiteness (or beta-function vanishing) condition
for nonlinear sigma model \cite{Tarb2}.

The sigma model action depends only on the metric
structure. Therefore it is surprising that the counterterms of the
sigma model with affine-metric manifold differ from counter terms
of the sigma model with Riemannian manifold \cite{Tarb1}. 
This difference can not be reduced to the metric redefenition
caused by infinitesimal coordinate transformation \cite{Alv} or to the 
nonlinear renormalization of the quantum fields \cite{Stel1}.
In the paper \cite{Tarb1}, the counter terms are calculated for
conventional sigma model without assuming a metric connection for the
geodesic line equation in covariant background field method. In this
approach the connection of the sigma model equation of motion is a
metric connection and we must use a manifolds with two different
connection structures and the metric structure. Therefore it seems
more natural for the nonlinear sigma model with nonmetric manifold
to consider both the geodesic line equation and the sigma model
equation with nonmetric connection (i.e. not derived from
the metric). It leads to a generalization of the usual sigma model
which describes the string propagation in affine-metric curved space.
String motion on the nonmetric (affine-metric) manifold can be
considered as the motion of the string subjected to the dissipative
forces. In order to see it we discuss a relationship between the
geometrical structures of the manifold and the equation of motion. 

The equation of motion for the particle subjected to the forces
$Q^i(q,u)$ has the form  
\be 
du^i /dt - Q^i(q,u) = 0 
\ee
where $q^i$ are the coordinates and  $ u^i = dq^i /dt \ ( i=1,...,n ) $ . 
We suggest that eq. (1) are invariant under general coordinate
transformations and that for simplicity $Q^i(q,u)$ are the
gomogeneous functions of second power of $u$.
It is known that the local Lagrange function exists
and eq. (1) can be derived from least action principle if
and only if the Helmholtz conditions are satisfied. In this case
there are matrix multipliers \cite{Hav2,Edv} such that eq. (1)
becomes Euler-Lagrange equation.
The spesial case  $Q^i(q,u) = - {[^i}_{kl}] u^k u^l$, where ${[^i}_{kl}]$ 
is a Christoffel symbol, the n-dimensional curved space is Riemannian
manifold and eq. (1) defines the usual one-dimensional
nonlinear sigma model. 
On the other hand it is known that
Lagrange function uniquely defines the metric structure on the
$(n+1)$- dimensional configurational space \cite{Lanc}. That is the
equation of motion derived from least action
principle is equivalent to the geodesic line equation on metric
manifold. The connection structure can be naturally defined on the metric
manifold as  Christoffel symbols. 
As the result the motion of the system subjected to
potential forces is equivalent to the free motion of the test
particle on the metric (Riemann, Finsler or Kawaguchi) manifold, i.e.
manifold which connection and metric structures are correlated.

If the Helmholtz conditions are not satisfied, the equation of motion (1) 
can be represented as the particle motion subjected to 
dissipative forces $Q_d^i$ on the metric manifold with metric structure
defined by the Lagrangian 
\be
\frac{d u^i}{dt} - Q_p^i (q,u) - Q_d^i (q,u) = - (g^{-1})^{ij}D_j
L(q,u) - Q_d^i (q,u) = 0 
\ee                
where $D_j$ is the Euler-Lagrange operator, $L(q,u)$ the Lagrange function
and $ g_{ij}(q,u)$ the matrix multiplier \cite{Hav2}. 
The dissipative force for the one-dimensional sigma model with affine-metric 
field manifold is defined by the connection defect 
$Q^i_d = - {D^i}_{kl}(q) u^k u^l$.
If the free motion of the test particle on the manifold are defined
by eq. (2) then this manifold is nonmetric.
This manifold usually called generalized path space \cite{Doug} 
and allow naturally to define connection structure which coefficients are
$ {\Gamma^i}_{kl} (q,u) = (-1/2) (\partial^2 Q^i / \partial u^k \partial
u^l) $.  
In the generalized path space the connection structure is not
correlated with the metric structure of this space.  
As the result  the motion of
the systems subjected to dissipative forces on the metric manifold
is equivalent to the free motion of the test particle on the
nonmetric (generalized path) manifold.
Note that the equation of motion
and the geodesic line equation in the nonmetrical manifold can be
derived from Sedov variational principle
\cite{Sed} which is the
generalization of the least action principle.

The affine-metric manifold \cite{Sch} (path space with metric
\cite{Veb}) is a simple example of the 
generalized path space with a metric structure.
That is the consistent approach to the nonlinear sigma model with
affine-metric  manifold lead to a generalization of the usual
one-dimensional sigma model which represents a particle  subject to
dissipative forces. 
Analogously we have that {\it the motion of the string in
affine-metric curved space is equivalent to the motion of the string
subjected to dissipative forces on  Riemannian manifold \cite{Tar4}.}
For this reason the consistent theory of the bosonic string in the
curved affine-metric space is a quantum dissipative theory. 
Note that the dissipative models in fundamental interactions theories are
discussed in \cite{Prig,Ell,Tar4}.  

The quantum description of the dissipative systems without well-known
ambiguities \cite{Lem,Hav2,Edv,Prig,Tar4}, without nonassociative
violation of the canonical commutation relations \cite{Sant1} and
beyond the sphere of quantum kinetics is suggested in
\cite{Tar2,Tar3,Tar4}. This description uses Sedov variational
principle in the phase space to generalize the canonical quantization. 
The suggested quantization does not violate Heisenberg algebra 
because it generalizes the canonical quantization by introducing the operator
of the nonholonomic quantities in addition to the usual associative
operators of the momentum, coordinate and holonomic functions.
The generalization of the von Neumann equation was derived from the 
dissipative Liouville equation \cite{Steb,Tar3} contrary to the usual heuristical 
and therefore ambiguous generalization \cite{Neum,Ell}.

In ref. \cite{Tar3}  the conformal anomaly of the energy
momentum tensor trace  for closed
bosonic string on the affine-metric manifold is considered and  it is  proved 
from the conformal invariance that metric and dilaton beta-functions
of the sigma model with affine-metric field manifold must be trivial 
as usual \cite{Shor}. 

In the present paper the two-loop ultraviolet
metric counterterms and beta-function  for
the two-dimensional nonlinear sigma model with affine-metric field
manifold are calculated. {\it The correlation between the connection and the metric
structures on the manifold are derived from  the beta-function
vanishing condition. }

\section{One-loop and two-loop calculations.}

Let us consider now the closed bosonic string in curved
space-time \cite{Lov}.  
The world sheet swept out by the string is described
by the map $ X(x) $ from two-dimensional parameter space N into
n-dimensional space-time manifold M, i.e. $X(x):$ $N \rightarrow M$.
The two-dimensional parameter is $ x = ( \tau, \sigma ) $ and the map
$X(x)$ is given by space-time coordinates $ X^k (x) $.
The classical equation of motion for the closed bosonic string in 
the n-dimensional affine-metric curved space-time has the form
\be
\partial_{\mu}  \sqrt{g} g^{\mu \nu} \partial_{\nu} X^i + 
{\Gamma^i}_{kl}(X) \partial_{\mu} X^k  
\sqrt{g} g^{\mu \nu} \partial_{\nu} X^l = 0 
\ee
where $ g^{\mu \nu} (x) $ is the two-dimensional metric tensor; 
$ \ {\Gamma^i}_{kl}(X)$ the affine connection, which can be represented in
the form $ {[^i}_{kl}] + {D^i}_{kl} \ $; 
$ \ {[^i}_{kl}] $ is the Christoffel symbol for the metric $ G_{ij} (X) $ ; 
$ \ D_{ikl} (X) $ is a connection defect tensor which can be 
written in the form \cite{Sch} 
\be
{D^i}_{kl} (X) = (-1/2) G^{ij} (K_{jlk} + K_{jkl} - K_{klj}) + 2
{Q_{(kl)}}^i + {Q^i}_{kl}
\ee
where $K_{kli} = \nabla_i G_{kl} $ is nonmetricity tensor and
${Q^i}_{kl} $ is torsion tensor.
 The equation of motion (3)  is an equation of the 
two-dimensional geodesic flow on the affine-metric manifold (the 
two-dimensional analogue of the geodesic line). It is well known that 
this equation can not be derived from the least action principle. 
Note that the 
Riemannian geodesic flow (${D^i}_{kl} = 0 $) can be derived from 
this variational principle 
with the Lagrangian defined by
%%if the holonomic functional (action) is defined by
\be
L(X) = (1/2) \ G_{kl} (X) \partial_{\mu} X^k \sqrt{g} 
 g^{\mu \nu} \partial_{\nu} X^l 
\ee 
The affine-metric geodesic flow equation (3) can be derived from 
the Sedov variational principle \cite{Sed}
 if the variation of the nonholonomic functional has the form 
\be
\delta \tilde W = \int d^2 x \ \delta  W = - \int d^2 x \ D_{ikl} (X) \ 
\partial_{\mu} X^k  \sqrt{g} 
g^{\mu \nu} \partial_{\nu} X^l \ \delta X^i  
\ee
The holonomic and nonholonomic functionals define a closed bosonic 
string propagating in the affine-metric curved space-time or in the
presence of dissipative and nondissipative background fields.

In loop calculation we use the generating functional for 
connected Green functions
in the phase-space path-integral form for non-Hamiltonian
(dissipative) systems
suggested in \cite{Tar2,Tar3,Tar4}. This  generating functional 
is written in the form 
\be
 Z (J,g) = - \imath  \ ln \int DX DP \ exp \ \imath 
\int d^2x \ ( P_k (d X^k / d \tau) - H + W + 
( \imath / 2 ) \Omega + K (J) ) 
\ee
where $K (J)$ is the source term; $\Omega$ is defined in the Appendix 
(eq. (17)) and $\hbar = 1$ . 
To perform the calculation of the on-shell ultraviolet behavior in
one- and two-loop order for sigma model we use the affine-metric
covariant background field expansion in normal coordinates  
\cite{Veb,Tarb1} and new generating functional  $ Z (X_0, g, J) $.
The covariant background field method \cite{Alv,Tar2} in the phase space
is defined by the usual  expansion of the coordinates $ X^k (x) $ only.
Note that the background field method can be considered as 
conservative model approximation for the quantum dissipative models.
The generating functional  $Z ( X_0, g, J )$ is defined by
\be
exp \ \imath Z ( X_0 , g , J )  =  \int D \xi D P \ exp \ \imath \int d^2 x 
\ ( P_k \frac{d} { d \tau}  X^k - H + W + 
\frac{\imath }{ 2} \Omega  + J_k \xi^k \ ) 
\ee
where $X = X(X_0,\xi)$ ; $X^i_0(x)$ is the solution of classical
equation of motion; $\xi^k(x)$ the covariant field which is the tangent
vector to the affine-metric geodesic line containing $ X_0^k $ and $ X^k $ . 

We produce the Hamiltonian, 
nonholonomic functional and omega function  in the conformal gauge as 
a power series in the field $ \xi^k (x) $ :
\be
H = \ - \ (1/2) \ G^{kl} (X) P_k P_l \ - \ (1/2) \  
G_{kl} (X) {X^{\prime}}^k {X^{\prime}}^l 
\ee
\be 
W \ =  \ (1/2) \ {\Delta_1}^{kl} P_k P_l \ +  
\ (1/2) \ {\Delta^2}_{kl} {X^{\prime}}^k {X^{\prime}}^l 
\ ; \quad \Omega = \ 2 \ D^k (X) P_k 
\ee
where $X^i = X^i (X_0,\xi)$; $ D^k (X) \equiv {D^k}_{ij} (X) \ G^{ij}
(X) $; $ {X^{\prime}}^k \equiv (d X^k)/(d \sigma ) $; $P_k $ is the
canonical momentum.  The background field expansions of the 
$\Delta$-operators are written in the form  
\be
{\Delta_1}^{kl} = 2 \ {D_i}^{kl} (X_0) \ \xi^i + \ O ( \xi^2 ) \ ; \quad 
{\Delta^2}_{kl} = - 2 \ D_{ikl} (X_0) \ \xi^i + \ O ( \xi^2 ) 
\ee

To obtain all of the one- and two-loop counterterms we need to expand
Lagrangian, non-holonomic functional and omega function to fourth
order in the quantum fields $\xi^a (x) $. 
The functional integral of  $Z(X_0,g,J)$ over momentum $ P $ is
the Gaussian integral. 
It is easy to derive the path integral form for the generating functional:
\be
Z (X_0, g, J) = - \imath  \ ln \int D \xi  \ exp \ \imath  
 \  \int d^2 x \ A (X (X_0, \xi))    
\ee
The full expression of $A(X)$ is complicated. Therefore let us consider 
terms of $A(X)$ which give the nontrivial simple poles two-loop
metric divergences only: 
\[ A(X_0, \xi) = (1/2)  \partial_{\mu} \xi^a \partial_{\mu} \xi^a 
+ A_{abk}  \partial_{\mu} X^k_0  \xi^a \partial_{\mu} \xi^b +  B_{abkl}
\xi^a \xi^b   \partial_{\mu} X^k_0  \partial_{\mu} X^l_0 +     
 J_{abc} \xi^a \partial_{\mu} \xi^b \partial_{\mu} \xi^c + \] 
\[ +  C_{abcl}  \partial_{\mu} X^l_0  \xi^a \xi^b \partial_{\mu} \xi^c    
 +  L_{abcd} \xi^a \xi^b \partial_{\mu} \xi^c \partial_{\mu} \xi^d 
+ E_{abcdp} \partial_{\mu} X^p_0 \xi^a \xi^b \xi^c \partial_{\mu} \xi^d + 
 F_{abcd} \xi^a \xi^b \partial_{\mu} \xi^c \kappa^{\mu \nu} 
  \partial_{\nu} \xi^d \]  
where \\

\noindent
$  A_{abk} = [ G_{kj;i} + D_{i(jk)} -(1/2) G_{ij;k} ] e^i_a e^j_b 
\ ; \quad  J_{abc} = [ (1/2) G_{jk;i} + (1/3)D_{i(jk)} ]  e^i_a e^j_b e^k_c $\\

\noindent
$ B_{abkl} = [(1/2) \hat R_{kijl} + (1/4) G_{kl;ij} + (1/8) G_{pi;k}
G_{pj;l} - (1/2) G_{pj;k} G_{lp;i} + $ 

$ + (1/2)D_{i(kl);j}) - (1/2)D_{i(lp)} G_{pj;k} ]  e^i_a e^j_b  $\\

\noindent
$ C_{abcl} = [ (2/3) \hat R_{(k/ij/l)} + (1/2)G_{kl;ij} - (1/2) (G_{pk;i}
+ (2/3)D_{i(pk)}) G_{pj;l}  + (2/3)D_{i(kl);j} ]  e^i_a e^j_b e^k_c  $\\

\noindent
$  E_{ijklp} = [ (5/36) G_{n(l/;i} \hat R_{njk/p)} + (1/4) 
\hat R_{lijp;k} + (1/6) G_{lp;ijk} - (1/4)D_{i(nk);j} G_{nl;p}  +  $

$ + (1/6) \hat R_{n(ij)p} D_{k(nl)}  + (1/6) \hat R_{n(ij)l} D_{k(np)} +
(1/4)D_{(i/lp;/jk)}] e^i_a e^j_b e^k_c  e^l_d  $ \\

\noindent
$ L_{abcd} = [ (1/6) \hat R_{kijl} + (1/4) G_{kl;ij} + (1/4)D_{ikl;j}]
e^i_a e^j_b e^k_c  e^l_d   ; $ \\

\noindent
$ F_{ijkl} = 2 D_{i(nk)} D_{j(nl)}  e^i_a e^j_b e^k_c  e^l_d .   $ \\

In the conformal gauge kappa tensor has the form 
 $ \ \kappa^{\mu \nu} = ( \kappa^{\tau \tau}, 
\ \kappa^{\tau \sigma}, \ \kappa^{\sigma \sigma}) = (- 1, 0,  0 ) $. 
We use the following notations 
\[ \hat {R^i}_{jkl} = {R^i}_{jkl} + 2 \hat \nabla_{[l/} {Q^i}_{j/k]}  
+ 2 {Q^n}_{j[k/} {Q^i}_{n/l]} \ ; \qquad
 {R^i}_{jkl} =  2 \partial_{[k/} {\Gamma^i}_{j/l]}  
+ 2 {\Gamma^n}_{j[l/} {\Gamma^i}_{n/k]}   \]
\[\hat \nabla_k A_i =  \nabla_k A_i + {Q^n}_{ki} A_n = \partial_k A^i -
{\Gamma^n}_{(ki)} A_n  = A_{i;k} \ ; \qquad
 G_{ij;k} = K_{ijk} + 2 Q_{(i/k/j)} \]
\[ B_{[n/m}T_{/k]l} = (1/2) (B_{nm}T_{kl} - B_{km}T_{nl} ) \ ; \qquad
 B_{(j/k/l)} = (1/2) ( B_{jkl} + B_{lkj}) \]
and ${\Gamma^i}_{(kl)}$ is the symmetry part of the affine connection.
The terms of $A(X_0,\xi)$ are usual \cite{Frid,Alv} if and only if 
both the nonmetricity tensor $K_{ijl}$ and the symmetry part of
torsion $Q_{(jk)i} $ are equal to zero.

Note that in the expression $A(X_0,\xi)$ we take into account the
additional nonmetric terms caused by the following. It is known that
propagator of the quantum fields $\xi^k (x) $ is 
not standard. Therefore we introduce an n-bein $ e^a_k (X) $ 
and define $\xi^a (x) = e^a_k \xi^k (x) $, where $ \hat \nabla_k e^a_l = 0$.
After this modification the kinetic terms become $ \hat \nabla_{\mu}
\xi^a \hat \nabla_{\mu} \xi^a $ , where 
$ \hat \nabla_{\mu} \xi^a = \partial_{\mu} \xi^a + \hat
{\Lambda^a}_{bc} \ e^b_k  \ \partial_{\mu} X^k_0  \ \xi^c $.
This mixed covariant derivative  for the affine-metric manifold M 
and the Minkowski space N involves the Schouten-Vranceanu 
connection \cite{Schot} $ \ \hat \Lambda_{abc} \ $,  which is 
equal to the Ricci rotation coefficient \cite{Chan} and the object 
$ \omega^a_{kc} \equiv \hat \Lambda^a_{bc} e^b_k $ is spin connection
\cite{Alv} on the Riemannian manifold. Note in addition to diagrams
of \cite{Tarb1} we take into account the diagrams whose external background
field lines involve the Schouten-Vranceanu connection. 
This diagrams must not cancel \cite{Tar3} in contrary to the 
usual nonlinear sigma model \cite{Alv} and give the tensor
contribution. It caused by the relation  
\[ \ \hat \Lambda_{(a/b/c)} = ( - 1/2) (K_{ijl} + 2 Q_{(i/l/j)})
 \ e^i_a e^j_c e^l_b . \] 

The irreducible one-loop diagrams (figs.1a, 1b) produce the following
simple poles divergences: \\

\noindent
$ (1a) =  - (\mu^{2 \varepsilon} /4 \pi \varepsilon) B_{aakl} 
   \partial_{\mu} X^k_0  \partial_{\mu} X^l_0 $ \\

\noindent
$ (1b) = (\mu^{2 \varepsilon}/8 \pi \varepsilon) A_{[ab]k} A_{[ab]l} 
   \partial_{\mu} X^k_0  \partial_{\mu} X^l_0 $ \\

{\footnotesize
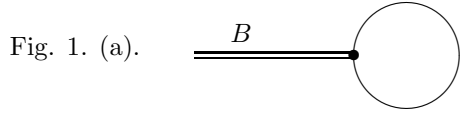
\begin{picture}(0,150) 
\put(0,100){Fig. 1. (a).}
\put(150,100){\circle{40}}
\put(130,100){\circle*{4}}
\put(70,101){\line(1,0){60}}
\put(70,99){\line(1,0){60}}
\put(80,105){ $ B $ }
\end{picture} 

\vskip -3cm
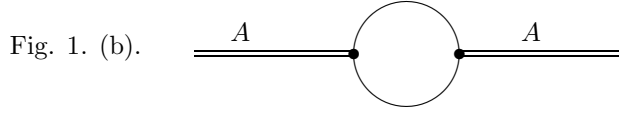
\begin{picture}(0,150) 
\put(0,100){Fig. 1. (b).}
\put(150,100){\circle{40}}
\put(130,100){\circle*{4}}
\put(170,100){\circle*{4}}
\put(70,101){\line(1,0){60}}
\put(70,99){\line(1,0){60}}
\put(170,101){\line(1,0){60}}
\put(170,99){\line(1,0){60}}
\put(80,105){ $ A $ }
\put(190,105){ $ A $ }
\end{picture} 
}
\vskip -2cm

The nontrivial simple poles ultraviolet two-loop divergences are
caused by the graphs of figs. 2-6. 
The two-loop simple poles divergences of these graphs are the following: \\

\noindent
$ (2a) = (\mu^{2 \varepsilon} /16 \pi^2 \varepsilon)
 C_{(ab)ck} C_{a[bc]l}  \partial_{\mu} X^k_0  
\partial_{\mu} X^l_0  $ \\

\noindent
$ (2b) = (\mu^{2 \varepsilon} / 16 \pi^2 \varepsilon)
 (J_{c(ab)} - J_{a(bc)}) (C_{(ab)cl;k}  +  D_{n(ka)}
C_{(nb)cl} +  D_{n(kb)} C_{(an)cl}  + $ 

$ +  D_{n(kc)} C_{(ab)nl} ) 
\partial_{\mu} X^k_0 \partial_{\mu} X^l_0  $\\

\noindent
$ (2c) = ( \mu^{2 \varepsilon} /32 \pi^2 \varepsilon)
   ( J_{a(bc);l}-J_{c(ab);l}  + D_{n(la)} J_{n(bc)} + D_{n(lb)}J_{a(nc)}
+ D_{n(lc)} J_{a(bn)} $

$ -  D_{n(lc)}J_{n(ba)} -   D_{n(lb)}J_{c(na)} - D_{n(la)}J_{c(bn)})
 ( J_{a(bc);l} + D_{n(ka)} J_{n(bc)} + D_{n(kb)}J_{a(nc)} + $

$ + D_{n(kc)} J_{a(bn)}) \partial_{\mu} X^k_0 \partial_{\mu} X^l_0 $\\

{\footnotesize

\begin{picture}(0,150) 
\put(0,50){Fig. 2 (a), (b), (c).}
\put(150,100){\circle{40}}
\put(130,100){\circle*{4}}
\put(170,100){\circle*{4}}
\put(70,101){\line(1,0){60}}
\put(70,99){\line(1,0){60}}
\put(170,101){\line(1,0){60}}
\put(170,99){\line(1,0){60}}
\put(130,100){\line(1,0){40}}
\put(70,105){ $ C,C,J $ }
\put(190,105){ $ C,J,J $ }
\end{picture} 
}

\noindent
$ (3a) =  - (\mu^{2 \varepsilon}/16 \pi^2 \varepsilon)
 ( L_{cc(ab)} + L_{(ab)cc}) B_{(ab)kl}  \partial_{\mu} X^k_0
 \partial_{\mu} X^l_0  $ \\

\noindent
$ (3b) = ( 3 \mu^{2 \varepsilon}/32 \pi^2 \varepsilon)
E_{(cca)bk} A_{[ab]l} \partial_{\mu} X^k_0 \partial_{\mu} X^l_0  $ \\

\noindent
$ (3c) = - (\mu^{2 \varepsilon}/32 \pi^2 \varepsilon)
 ( L_{cc(ab)} + L_{(ab)cc}) 
(A_{(ab)k;l}+D_{n(ka)}A_{nbl} + D_{n(kb)}A_{anl}) \partial_{\mu} X^k_0
\partial_{\mu} X^l_0  $ \\

\noindent
$ (3d) = (\mu^{2 \varepsilon}/16 \pi^2 \varepsilon)
 ( (-1/2) F_{cc(ab)} + (f_1+(1/2))F_{(ab)cc}) B_{(ab)kl}
\partial_{\mu} X^k_0 \partial_{\mu} X^l_0   $ \\

\noindent
$ (3e) = (\mu^{2 \varepsilon}/32 \pi^2 \varepsilon)
 ( (f_1+1/2) F_{cc(ab)}  - (1/2) F_{(ab)cc})
( A_{(ab)k;l} + D_{n(ka)} A_{nbl} + $ 

$ +  D_{n(kb)}A_{anl})  \partial_{\mu} X^k_0  \partial_{\mu} X^l_0  $ \\

{\footnotesize

%\begin{picture}(0,150) 
%\put(0,50){Fig. 2 (a), (b), (c).}
%\put(150,100){\circle{40}}
%\put(130,100){\circle*{4}}
%\put(170,100){\circle*{4}}
%\put(70,101){\line(1,0){60}}
%\put(70,99){\line(1,0){60}}
%\put(170,101){\line(1,0){60}}
%\put(170,99){\line(1,0){60}}
%\put(130,100){\line(1,0){40}}
%\put(70,105){ $ C,C,J $ }
%\put(190,105){ $ C,J,J $ }
%\end{picture} 

%\vskip -2cm
\begin{picture}(0,150) 
\put(0,30){Fig. 3 (a), (b), (c), (d), (e).}
\put(150,100){\circle{40}}
\put(150,60){\circle{40}}
\put(150,80){\circle*{4}}
\put(170,60){\circle*{4}}
\put(90,81){\line(1,0){60}}
\put(90,79){\line(1,0){60}}
\put(170,61){\line(1,0){60}}
\put(170,59){\line(1,0){60}}
\put(60,85){ $ L,E,L,F,F $ }
\put(190,65){ $ B,A,A,B,A $ }
\end{picture} 
}

\noindent
$ (4a) = - (\mu^{2 \varepsilon}/8 \pi^2 \varepsilon)
L_{(ab)(cd)} A_{[ac]k} A^{\mu}_{[bd]l}  \partial_{\mu}
X^k_0 \partial_{\mu} X^l_0  $ \\

\noindent
$ (4b) = (\mu^{2 \varepsilon}/48 \pi^2 \varepsilon)
 ( L_{(ac)dd} - 2 L_{dd(ac)}) A_{[ab]k}
A_{[cb]l}  \partial_{\mu} X^k_0 \partial_{\mu} X^l_0   $ \\

\noindent
$ (4c,d) = (\mu^{2 \varepsilon}/ 32 \pi^2 \varepsilon)
 ((1/3) A_{[ab]k} F_{(ac)dd} - (f_1+(4/3)) F_{dd(ac)}
A^{\mu}_{[cd]k}) A_{[ab]l} \partial_{\mu}
X^k_0 \partial_{\mu} X^l_0   $ \\

{\footnotesize
\begin{picture}(0,150) 
\put(0,50){Fig. 4 (a), (c).}
\put(150,100){\circle{40}}
\put(150,60){\circle{40}}
\put(150,80){\circle*{4}}
\put(170,60){\circle*{4}}
\put(170,100){\circle*{4}}
\put(90,81){\line(1,0){60}}
\put(90,79){\line(1,0){60}}
\put(170,61){\line(1,0){60}}
\put(170,59){\line(1,0){60}}
\put(170,101){\line(1,0){60}}
\put(170,99){\line(1,0){60}}
\put(90,85){ $ L,F $ }
\put(190,65){ $ A,A $ }
\put(190,105){ $ A,A $ }
\end{picture} 

\vskip -2 cm
\begin{picture}(0,150) 
\put(0,50){Fig. 4. (b), (d).}
\put(150,100){\circle{40}}
\put(150,60){\circle{40}}
\put(150,80){\circle*{4}}
\put(130,100){\circle*{4}}
\put(170,100){\circle*{4}}
\put(90,81){\line(1,0){60}}
\put(90,79){\line(1,0){60}}
\put(130,101){\line(-1,0){60}}
\put(130,99){\line(-1,0){60}}
\put(170,101){\line(1,0){60}}
\put(170,99){\line(1,0){60}}
\put(90,85){ $ L,F $ }
\put(80,105){ $ A,A $ }
\put(190,105){ $ A,A $ }
\end{picture} 
}

\noindent
$ (5a) = (\mu^{2 \varepsilon}/ 16 \pi^2 \varepsilon)
 ((3/2) J_{d(bc)} B_{adkl} +  J_{a(bd)} B_{cdkl} - 
2 J_{b(ad)} B_{cdkl} + $ 

$ + 2 J_{d(ac)} B_{bdkl}) J_{a(bc)} \partial_{\mu} X^k_0
\partial_{\mu} X^l_0 $ \\

\noindent
$ (5b) = (\mu^{2 \varepsilon}/16 \pi^2 \varepsilon)
 (2 J_{d(bc)} A_{[ad]k} +  (-2) J_{b(cd)} A_{[ad]k} +
2 J_{a(bd)} A_{cdk} ) C_{(ab)cl}  \partial_{\mu} X^k_0
\partial_{\mu} X^l_0 $ \\

\noindent
$ (5c) = (\mu^{2 \varepsilon}/32 \pi^2 \varepsilon)
 ((3/2) J_{abd} J_{cbd} +  J_{bda} J_{bdc} +
(-2) J_{bda} J_{dbc}) ( A_{(ab)k;l} + D_{n(ka)}A_{ncl} + $

$ + D_{n(kc)}A_{anl} ) \partial_{\mu} X^k_0  \partial_{\mu} X^l_0  $ \\

{\footnotesize
\begin{picture}(0,150) 
\put(0,40){Fig. 5. (a), (b), (c).}
\put(150,100){\circle{40}}
\put(130,100){\circle*{4}}
\put(170,100){\circle*{4}}
\put(150,80){\circle*{4}}
\put(70,101){\line(1,0){60}}
\put(70,99){\line(1,0){60}}
\put(170,101){\line(1,0){60}}
\put(170,99){\line(1,0){60}}
\put(151,80){\line(0,-1){40}}
\put(149,80){\line(0,-1){40}}
\put(130,100){\line(1,0){40}}
\put(80,105){ $ J,J,J $ }
\put(190,105){ $ J,C,J $ }
\put(155,50){ $ B,A,A $ }
\end{picture} 
}

\noindent
$ (6a) =  (\mu^{2 \varepsilon} /16 \pi^2 \varepsilon)
 ( (5/3) J_{abd} J_{cbd} - 
 (28/3)  J_{bda}  J_{cbd}  -  4  J_{bda}  J_{dbc} + $ 

$ +  6 J_{bda} J_{bdc} ) 
A_{[as]k} A_{[cs]l} \partial_{\mu} X^k_0 \partial_{\mu} X^l_0  $ \\

\noindent
$ (6b) = (\mu^{2 \varepsilon}/16 \pi^2 \varepsilon)
 (J_{bap} J_{dcp}   -  2 J_{pab} J_{dcp}  +  
(-1/2) J_{pab} J_{pcd} + $ 

$ + 2 J_{abp} J_{dcp} ) 
A_{[ca]k} A_{[bd]l} \partial_{\mu} X^k_0 \partial_{\mu} X^l_0 $ \\

where $B_{ac;c}T_{a}= B_{ac;d}T_{b}G^{ab}G^{cd}$.\\

\vskip 1cm
{\footnotesize
\begin{picture}(0,150) 
\put(0,50){Fig. 6. (a).}
\put(150,100){\circle{40}}
\put(130,100){\circle*{4}}
\put(170,100){\circle*{4}}
\put(150,80){\circle*{4}}
\put(150,120){\circle*{4}}
\put(70,101){\line(1,0){60}}
\put(70,99){\line(1,0){60}}
\put(170,101){\line(1,0){60}}
\put(170,99){\line(1,0){60}}
\put(151,80){\line(0,-1){40}}
\put(149,80){\line(0,-1){40}}
\put(151,120){\line(0,1){40}}
\put(149,120){\line(0,1){40}}
\put(150,80){\line(0,1){40}}
\put(90,105){ $ A $ }
\put(190,105){ $ A $ }
\put(155,60){ $ J $ }
\put(155,130){ $ J $ }
\end{picture} 

\begin{picture}(0,150) 
\put(0,50){Fig. 6. (b).}
\put(150,100){\circle{40}}
\put(130,100){\circle*{4}}
\put(170,100){\circle*{4}}
\put(150,80){\circle*{4}}
\put(150,120){\circle*{4}}
\put(70,101){\line(1,0){60}}
\put(70,99){\line(1,0){60}}
\put(170,101){\line(1,0){60}}
\put(170,99){\line(1,0){60}}
\put(151,80){\line(0,-1){40}}
\put(149,80){\line(0,-1){40}}
\put(151,120){\line(0,1){40}}
\put(149,120){\line(0,1){40}}
\put(150,80){\line(1,1){21}}
\put(90,105){ $ A $ }
\put(190,105){ $ J $ }
\put(155,60){ $ J $ }
\put(155,130){ $ A $ }
\end{picture} 
}

The divergent integrals are calculated using the dimensional
regularization (in $ n= 2 - 2 \varepsilon$ dimensions) with minimal
subtraction and  general prescription
for contraction of the two-dimensional $\kappa^{\mu \nu}$ tensor \cite{Tar3}
$ \kappa^{\mu \nu} \eta_{\mu \nu} = f (n) $ where 
$ f(n) = 1 + f_1 \varepsilon + O( \varepsilon^2 ) $ and  $ \eta_{\mu \nu} $ 
is two-dimensional Minkowski metric. The different  prescriptions 
may correspond to the different renormalization schemes and thus
their results should be related through redefenition of the couplings
by analogy to the two-dimensional nonlinear
sigma-model with  Wess-Zumino term \cite{Mets}. 
To distinguish between infrared and ultraviolet divergences we
introduce an  auxilliary mass term \cite{Tsey}.

The two-loop simple poles divergences caused by one-loop counterterms are
derivable from 
\be
 \Delta L^{(1)} =  \frac{\mu^{2 \varepsilon}}{4 \pi \varepsilon}
 (P_{ab} \partial_{\mu} \xi^a \partial_{\mu} \xi^b
+ V_{abk} \partial_{\mu} X^k_0  \xi^a \partial_{\mu} \xi^b  +   
\mu^2  M_{ab} \xi^a \xi^b ) , 
\ee
where \\  

\noindent
$ P_{ab} = [- B_{ccij} + (1/2) A_{[cd]i} A_{[cd]j}]  e^i_a e^j_b  $\\

\noindent
$ V_{abk} = [ 2 P_{kj;i} - P_{nj} G_{ni;k} ]  e^i_a e^j_b $; \\ 

\noindent
$M_{ab} = [ (-1/6) \hat R_{(i/nn/j)} - G_{in;nj} - G_{ik;l} G_{kl;j} ]
 e^i_a e^j_b . $ \\

The simple poles divergent part of the graphs (figs.7, 8) are \\

\noindent
$ (7a) = - ( \mu^{2 \varepsilon}/32 \pi^2 \varepsilon)
 P_{(ab)}  (A_{(ab)k;l}+D_{n(ka)}A_{ncl} +
D_{n(kc)}A_{anl})  \partial_{\mu} X^k_0  \partial_{\mu} X^l_0 $\\

\noindent
$ (7b) = - (\mu^{2 \varepsilon}/16 \pi^2 \varepsilon)
 P_{(ab)} B_{(ab)kl} \partial_{\mu} X^k_0  \partial_{\mu} X^l_0  $\\

\noindent
$ (7c) = (\mu^{2 \varepsilon}/32 \pi^2 \varepsilon)
V_{[ab]k} A_{[ab]l} \partial_{\mu} X^k_0
\partial_{\mu} X^l_0 $ \\

{\footnotesize
\begin{picture}(0,150) 
\put(0,50){Fig. 7. (a), (b), (c).}
\put(150,100){\circle{40}}
\put(130,100){\circle{13}}
\put(170,100){\circle*{4}}
\put(125,95){\line(1,1){10}}
\put(125,105){\line(1,-1){10}}
\put(70,101){\line(1,0){60}}
\put(70,99){\line(1,0){60}}
\put(170,101){\line(1,0){60}}
\put(170,99){\line(1,0){60}}
\put(73,110){ $  P, P, V$ }
\put(175,110){ $ A, B, A $ }
\end{picture} }

\noindent
$ (7d) = (  2 \mu^{2 \varepsilon}/48 \pi^2 \varepsilon)
P_{(ab)} A_{[ac]k} A_{[bc]l}  \partial_{\mu}
X^k_0  \partial_{\mu} X^l_0  $\\

{\footnotesize
\begin{picture}(0,150) 
\put(0,50){Fig. 7. (d).}
\put(150,100){\circle{40}}
\put(150,80){\circle{13}}
\put(170,100){\circle*{4}}
\put(130,100){\circle*{4}}
\put(145,75){\line(1,1){10}}
\put(145,85){\line(1,-1){10}}
\put(70,101){\line(1,0){60}}
\put(70,99){\line(1,0){60}}
\put(149,80){\line(0,-1){40}}
\put(151,80){\line(0,-1){40}}
\put(170,101){\line(1,0){60}}
\put(170,99){\line(1,0){60}}
\put(80,110){ $  A $ }
\put(190,110){ $ A $ }
\put(155,60){ $ P $ }
\end{picture} 
}

\noindent
$ (8a) =  (\mu^{2 \varepsilon}/16 \pi^2 \varepsilon)
M_{(ab)} B_{(ab)kl} \partial_{\mu} X^k_0  \partial_{\mu} X^l_0 $ \\

\noindent
$ (8b) =  (\mu^{2 \varepsilon}/32 \pi^2 \varepsilon)
M_{(ab)}  (A_{(ab)k;l}+D_{n(ka)}A_{ncl} +
D_{n(kc)}A_{anl})  \partial_{\mu} X^k_0  \partial_{\mu} X^l_0  $ \\

{\footnotesize
\begin{picture}(0,150) 
\put(0,50){Fig. 8. (a), (b).}
\put(150,100){\circle{40}}
\put(130,100){\circle{13}}
\put(170,100){\circle*{4}}
\put(125,95){\line(1,1){10}}
\put(125,105){\line(1,-1){10}}
\put(70,101){\line(1,0){60}}
\put(70,99){\line(1,0){60}}
\put(170,101){\line(1,0){60}}
\put(170,99){\line(1,0){60}}
\put(73,110){ $  M, M $ }
\put(175,110){ $ B, A $ }
\end{picture} 
}

\noindent
$ (8c) = - (\mu^{2 \varepsilon}/48 \pi^2 \varepsilon)
M_{(ab)} A_{[ac]k} A_{[bc]l}  \partial_{\mu}
X^k_0  \partial_{\mu} X^l_0  $ \\

{\footnotesize
%\vskip -2 cm
\begin{picture}(0,150) 
\put(0,50){Fig. 8. (c).}
\put(150,100){\circle{40}}
\put(150,80){\circle{13}}
\put(170,100){\circle*{4}}
\put(130,100){\circle*{4}}
\put(145,75){\line(1,1){10}}
\put(145,85){\line(1,-1){10}}
\put(70,101){\line(1,0){60}}
\put(70,99){\line(1,0){60}}
\put(149,80){\line(0,-1){40}}
\put(151,80){\line(0,-1){40}}
\put(170,101){\line(1,0){60}}
\put(170,99){\line(1,0){60}}
\put(80,110){ $ A $ }
\put(190,110){ $ A $ }
\put(155,60){ $ M $ }
\end{picture} 
}
 
The full expression for the metric beta-function is complicated. Let us
consider the special form of the nonmetricity tensor:
$ K_{ijl} = N_{ijl}= N_{(ijl)} $, where $ Q_{(ij)l} = 0 $  and 
$ N_{ij(l;k)} = {N^n}_{i(k} N_{l)jn} $. 
The two-loop metric beta-function \cite{Frid} for the 
bosonic nonlinear two-dimensional sigma model with this affine-metric
field manifold has the form 
\[ \beta^G_{kl} = (1/2 \pi) 
[ (1/8)  N_{nm(k} N_{l)nm} - (1/2) \hat R_{(k/nn/l)} ]  +   
(1/4 \pi^2) [ (1/2) ((2/3) \hat R_{(c/(ab)/k)} - \] 
\[- (1/6) N_{n(c/(a} N_{b)/k)n}) 
 ((2/3) \hat R_{(c/(ab)/l)} - (2/3) \hat R_{(b/(ac)/k)} + 
 (1/6) N_{n(b/(a} N_{c)/k)n}  - \] 
\[ - (1/6) N_{n(c/(a} N_{b)/l)n} ) +   ((1/2) \hat R_{(k/(ab)/l)} - 
 (1/8) N_{n(a/(k} N_{l)/b)n} ) ((1/6) \hat R_{(a/nn/b)}  - \]
\be
- (1/6) \hat R_{n/(ab)/n)} - ((151/72) + (1/2) f_1) N_{nm(a} N_{b)nm}) ] 
\ee
This metric beta-function leads to the well-known equation
\cite{Frid,Alv} on the Riemannian manifold ($K_{ijl} = 0$ and 
${Q^i}_{kl} = 0$).

It is easy to see the following ultraviolet finiteness conditions. 
The one loop and two loop parts of the
metric beta-function  for the two-dimensional nonlinear sigma model 
with affine-metric manifold 
vanish if the correlation between the affine connection and the 
metric structures on the manifold M is given by \\

\noindent
$  \nabla_{l} G_{ij}  = N_{ijl} = N_{(ijl)} ;$\\ 

\noindent
$Q_{(ij)l} = 0 \ ;$ \\

\noindent
$\hat \nabla_{(l} N_{k) ij} = {N^p}_{i(k} N_{l)jp} \ ;$ \\ 

\noindent
$\hat R_{(k/(ij)/l)} =  (1/4) \ {N^p}_{(k/(i} N_{j)/l) p} .$\\

These conditions have not the $ f_1 $ dependence and
define nonflat space,
i.e. the Riemannian curvature tensor is not equal to zero.
Note that the part of the metric beta-function 
from the sigma model action  only is zero in all 
loops if the affine-metric manifold  is defined by

\[ \hat R_{kijl} \ \equiv \ R_{kijl} - 2 \hat \nabla_{[j/} Q_{ki/l]} - 
2 {Q^n}_{i[l/} Q_{kn/j]} \ = \ 0  \ ; \] 

\[\hat \nabla_k G_{ij} \ = \ K_{ijk} - 2 Q_{(ij)k} \ = 0  \]

It is easy to see that this affine-metric manifold is not flat.\\

\section{Acknowlegment} 

I would like to thank Belokurov V.V. and Stelle K.S. for helpful and valuable 
discussions and Theoretical High Energy Physics Department of Nuclear Physics 
Institute of Moscow State University  for their support during the work.

\newpage

\section{Appendix}

The equation of motion and the geodesic line equation in nonmetrical
manifold can be derived from the Sedov variational principle
\cite{Sed} which is the generalization of the
least action principle: 
\be
\delta S(q) + \delta  \tilde W (q) = 0 
\ee
where S(q) is the holonomic functional called action and 
$  \tilde W (q) $ is the nonholonomic functional, i.e.  
$$ \  \delta \delta^{\prime} \tilde W \not\equiv  \delta^{\prime} \delta \tilde W .$$ 
For eq. (2) the nonholonomic functional has the form
\be
\delta \tilde W = \int d t \ \delta W = \int d t \ Q^i_d (q,u)
\ g_{ij} \ \delta q^j  
\ee
i.e. nonholonomic functional is defined by the connection defect.
Nonholonomic functional $W$ is characterized 
by the following properties in the phase space:  \\

\noindent
(1) \ $ [W, p_k] = W^q_k $ \ and \ $[W, q^k] = - W_p^k $ \
i.e. the variation of the functional $W$ is defined by 
$\delta W = W_k^q \delta q^k + W_p^k \delta p_k .$ 
The brackets are  the
generalized (variational) Poisson brackets \cite{Tar3,Tar4} which are 
coincide with usual Poisson brackets for the holonomic functions.  \\

\noindent
$ (2) \  J[Z_k,W,Z_l] = J_{kl} \not\equiv 0$ \ if \ $ k  \not\equiv l $\
where 
$J[A,B,C] = [A[BC]] + [B[CA]] + [C[AB]] ; \ k = 1,...,2n \ $
and $ Z_i = q^i \ $ and $ Z_{n+i} = p_i  \ $ if  $ i=1,..,n $.
The Jacobian $J_{kl}$ characterizes the deviation from the condition
of integrability. The object $W$ is the nonholonomic object if one of
the $J_{kl}$ is not trivial. \\

Note in addition that the classical phase space equation of motion
for dissipative systems has the form 
$ d Z_k / dt = [Z_k, H-W] $ and Liouville equation for dissipative systems
\cite{Steb,Tar3} has the form 
\be
\frac{d}{dt} \rho (q,p,t) = - \Omega (q,p) \ \rho (q,p,t)
\ , \ where \quad
\Omega (q,p) = \sum^n_{i=1} \ J[q^i,W,p_i] 
\ee

\end{document}